# PRINCIPAL COMPONENT ANALYSIS AND SELF ORGANIZING MAP FOR VISUAL CLUSTERING OF MACHINE-PART CELL FORMATION IN CELLULAR MANUFACTURING SYSTEM

Manojit Chattopadhyay

*Department of Computer Application, Pailan College of Management & Technology, Kolkata-7001041, West Bengal, India  email: chattomanojit@yahoo.com,*

Pranab K. Dan

*School of Engineering & Technology, West Bengal University of Technology, Kolkata 700 064, West Bengal, India email: dan1pk@hotmail.com*

Sitanath Majumdar

*Faculty Counsel for PG Studies in Commerce, Social Welfare & Business Management, Calcutta University, Alipore Campus, Reformatory Street, Kolkata - 700 027, West Bengal, India
email:  sitanath_mazumdar@rediffmail.com*

## Abstract

The present paper attempts to generate visual clustering and data extraction of cell formation problem using both principal component analysis (PCA) and self organizing map (SOM) from input of sequence based machine-part incidence matrix. First, the focus is to utilize PCA for extracting high dimensionality of input variables and project the dataset onto a 2-D space. Second, the unsupervised competitive learning of SOM algorithm is used for data visualization and subsequently, to solve cell formation problem based on ordinal sequence data via the node cluster on the SOM map. Although the numerically illustrated results from dataset revealed that PCA has explained most of the cumulative variance of data but in reality when the very large dimensional cell formation problem based on sequence is available then to obtain the clustering structure from PCA projection is become very difficult. Most importantly, in the visual clustering of ordinal data, the use of U-matrix alone can not be efficient to get the cluster structure but with color extraction, hit map, labeling via the SOM node map it becomes a powerful clustering visualization methodology and thus the present research contribute significantly in the research of cellular manufacturing.

## 1. Introduction

The primary concern in cellular manufacturing is utilization of group technology to identify similarities in processing requirements, operation sequence and other related production factors of a family of products to identify machines that can be clustered together in a cell so that intercellular movements of parts can be minimized. Thus the part family to be processed in a machine cluster will have similar operation sequence or other characteristics so that the overall efficiency of the manufacturing system will be enhanced. There are typically two approaches to consider the type of input for the above mentioned cell formation problem: first one is the most frequently used binary machine-part incidence matrix and second one is the non-binary i.e., operation sequence (ordinal data), operation time (ratio data). The machine-part incidence matrix $A = \{\{a_{ij}\}\}$ is derived from production route sheet. Where $a_{ij}$ means: '1' if parts i visits machine j otherwise '0'. This becomes input to any clustering algorithm so that family of parts to be processed in a cluster of machines can be identified from the output block diagonal matrix. In the block diagonal matrix if any production information like '1's appear outside cell represents inter-cell movement of parts (exceptional elements) and presence of any zero inside the cell represents as void. The perfect cell formation is achieved when the cells in diagonal contains all '1's. The principal objective of cellular manufacturing is to minimizing total material handling cost caused by inter-cellular moves, maximizing intra-cellular utilization of machines, and minimizing the duplication of machines in a GT cell (Gunasekharan et al, 1994). The sequence of operations in which parts are processed in machines represented by the ordinal numbers like 1,2,3 and so on otherwise non visit of parts to the machines are represented by zero. This ordinal number replaced the '1's in the cell formation problem and thus the matrix is represented as sequence based machine-part incidence matrix.

In literature there are numerous clustering algorithms applied to generate cell formation solution (Papaioannou and Wilson, 2010; Venkumar and Haq, 2006) viz., production flow analysis (Burbidge, 1963; Burbidge, 1971), similarity coefficient (Pollalis and Mavrommatis, 2009; Oliveira *et al.,* 2008; Yin and Yasuda, 2006), array based manipulation (Chan and Milner, 1982), graph theoretic (Mukhopadhyay *et al.,* 2009), clustering (Ünler and Güngör, 2009; Rogers and Kulkarni, 2005; George *et al.,* 2003; Nair and Narendran, 1998; Dimopoulos, 2006), mathematical programming (Mahdavi *et al*., 2010; Defersha and Chen, 2006; Choobineh, 1988), search method (Baykasoglu and Gindy, 2000; Lee-Post, 2000; Zhao and Wu, 2000), neural network (Ateme-Nguema and Dao, 2009; Pandian and Mahapatra, 2009; Venugopal and Narendran,1994; Soleymanpour *et al.,* 2002; Çakar and Cil, 2004), metaheuristic algorithm (Lin *et al.,* 2010; Li *et al.,* 2010; Venkataramanaiah, 2008; 1, Andrés and Lozano, 2006) etc.

Most of the research reported on clustering of the machine-part cell formation based on binary data without considering the operation sequence based machine-part incidence matrix. After reviewing related works on operation sequence based cell formation solution (Pandian and Mahapatra, 2009; Bu et al., 2009; Mahdavi and Mahadevan, 2008; Pandian and Mahapatra, 2008; Jayaswal and Adil, 2004; Ohta and Nakamura, 2002; Yin and Yasuda, 2002; Nair and Narendran 1998; Kang and Wemmerlov 1993; Harhalakis *et al.,*1990; Vakharia and Wemmerlov 1990; Choobineh 1988; Selvam and Balasubramanian 1985), the motivation of the present work to consider clustering of sequence based cell formation problem is due to the following reasons:

- it generates benefits of cellular manufacturing with less backtracking, reduced material handling, better control of cell activities, reduction in work-in-process inventory and lead time

- the objective of minimizing exceptional elements without considering operation sequences, does not lead to minimization of material handling as reported by Harhalakis *et al.,* (1990)
- the similarity of operation sequence in each cell facilitates the implementation of just-in-time manufacturing system
- non requirement of converted precedence incidence matrix from the sequence based machine-part incidence matrix and thus less time and computational burden (Won and Currie, 2007)
- as the part by part matrix for each machine generated from sequence based matrix itself becomes a large input matrix to the earlier clustering algorithm and direct input of sequence based machine-part matrix of industry size can be helpful to solve real life problems
- Venkumar and Haq (2006) reported SOM is suitable for cell formation solution to any size of machine-part incidence matrix

Due to the ready representation and interpretation, visualization and classification can efficiently handle the complexity of data as a standard tool. The cell formation is NP hard problem. In a large dimensional problem, a multivariate principal component analysis (PCA) approach can be used to reduce the dimension before the output factors are becoming input in a clustering routine (Kiang *et al.,* 2004). In the present paper, PCA a multivariate, statistical data analysis tool is used to extract the size dimension information after constructing a linear projection of the machine-part dataset in the 2-D plane of the map. PCA is powerful technique to extract data (Kwan *et al.,* 2001) and effective for exploratory analysis of data (Koua, 2003). Tipping and Bishop (1999) reported that PCA can be used for processing and visualizing data. Yeung and Ruzzo (2001) reported that principal component has varying degree of effectiveness in capturing cluster structure. They also have studied that first few principal components do not necessarily capture most of the cluster structure. On the other hand Laitinen *et al.,* (2002) reported that sometimes the visualization of PCA is not suitable for the complex dataset. In this case, SOM algorithm implemented through som toolbox can be an alternative method for optimal cluster visualization of those complex datasets (Chattopadhyay *et al.,* 2010).

The present paper attempts to generate visual clustering and data extraction of machine-part cell formation using both principal component analysis and self organizing map from input of ordinal data based machine-part incidence matrix. First, the focus is to utilize PCA for extracting high dimensionality of input machine-part variables and project the dataset onto a 2-D space (e.g., Verleysen, 2003; Annas *et al*, 2007 and many others). Second, the unsupervised competitive learning of SOM algorithm is used for data visualization and subsequently, to solve the machine-part cell formation problem based on ordinal sequence data via the node cluster on the SOM map.

Multivariate correlation and PCA approaches have been reported to apply as machine-part clustering tool to solve cell formation problem based on binary data (Arvindh and, Irani, 1991; Hachicha *et al*., 2006; Hachicha *et al*., 2008). Self Organizing Map (SOM) an artificial neural network using unsupervised learning network reported to use in cell formation solution (Chattopadhyay *et al.,* 2010; Venkumara and Haq, 2006; Ampazis and Minis, 2004; Inho and Jongtae, 1997; Kiang *et al*., 1995; Kulkarni and Kiang, 1995;

Venugopal and Narendran, 1994; Chu, 1993). In the literature of cellular manufacturing very few researchers reported the use of multivariate statistics of PCA and SOM visualization to the clustering of cell formation problem based on binary as well as non-binary data.

The proposed approaches differ from the earlier reported works:
i) both PCA and SOM algorithm applied on non-binary operation sequence based data projected based on visual clustering pattern
ii) apriori cluster number is not required in both PCA and SOM method
iii) the input matrix is not required to convert in any other form and the ordinal data based machine-part incidence matrix is becoming direct input to both algorithms
iv) both methods are flexible and robust in generating cell formation solution
v) interpretation of visual clustering of machine-part cell formation through in depth analysis of machine and part cluster structure
vi) large industry size input of sequence based cell formation problem as a solution in reality.

Thus the purpose of the present paper is to i) visually clustering of machine-part cell formation using both principal component analysis and self organizing map from ordinal sequence data (non-binary) based machine-part incidence matrix; ii) developed a comparative analysis of visual clustering using PCA and SOM and thereby establishing a better ordinal sequence data based cell formation approach iii) compare the performance in cell formation using PCA and SOM.

## 2. Proposed Methodology for machine-part Clustering

### 2.1. *Principal Component Analysis*

The Principal component Analysis (PCA) perhaps is the best known and oldest technique in multivariate analysis (Jolliffe, 1986; Preisendorfer, 1988). Pearson, 1901, was first to introduce it in the context to recast linear regression analysis into a new form. Thereafter, it was developed by Hotelling, 1933 in the field of psychometry. In the various real world problems the PCA is frequently used in the data set with some intrinsic complexity (Tuncer *et al.,* 2008 ; Horenko *et al.,* 2006; Rothenberger *et al.,*2003; Atsma and Hodgson, 1999; Barbieri *et al.,* 1999). The PCA is applied as a cluster analysis tool to form machine groups and part families (Hachicha *et al.,* 2006) simultaneously. Application of PCA has been in representing the data using a smaller number of variables (Wall *et al.,* 2002).

Using the PCA method we consider the machine-part incidence data into a matrix $A = {}_nX_m$, where the n –rows are associated with the parts and m-columns with the input (machine) variables. If we consider $\bar{x}_k$ is mean of m (machine) variables in the matrix A, the covariance matrix is given by

$$\Phi = -\sum \left( \phantom{x} - \overline{\phantom{x}} \right)\left( \phantom{x} - \overline{\phantom{x}} \right) \neq$$

Where j=1, 2... n, k=1, 2... m. There are two main steps to find a few orthogonal features, called principal components (PCs) of the matrix A, as follows:

Step1. Calculate the eigenvenctors and eigenvalues of the covariance matrix. Since the covariance matrix from equation (1) is square, a set eigenvalue $(\lambda_1, \lambda_2,...,\lambda_m)$ can be found by solving the determinant equation $|\Phi - \lambda I| = 0$. A set nonzero eigenvector $E = [e_1, e_2,...,e_m]$, corresponding to the relevant eigenvalue is found by using $|\Phi - \lambda I| e_i = 0, i = 1, 2,..., m$. Then a diagonal non zero eigenvalues matrix $(\wedge)$ can be constructed from the sorted eigenvalues $\lambda_1 \geq \lambda_2 \geq ... \geq \lambda_m$.

Step 2. Find the PCs of the covariance matrix $\Phi$ that can be generated using a process of singular value decomposition, which is given by $\Phi = E \wedge E^T$ \hfill (2)

The set of PCs is then represented as a linear combination of the original variables of

$$PC_m = e_m^T x.$$

The coefficient of values of the PCs could be used to identify similarity of parts and machines based on their operation requirement and sequence. Further, the identification of family of parts to be processed by cluster of machines can be interpreted using a PCs projection developed for this purpose. Johnson and Wichren (1998) reported that the first two PCs are usually used to project a 2-D plane of data if they provide at least 80% of cumulative variance. The percentage of variance (PV) for the PCs can be calculated by

$$PV = \frac{\lambda_k}{\lambda_1 + \lambda_2 + ... + \lambda_m} \times 100\% \hspace{1cm} (3)$$

The PCA from the perspective of statistical pattern recognition is having the practical value as it is an effective technique for dimension reduction (Haykin, 2008). The other main advantage of PCA is that once these patterns in the data are found, the data can be compressed by reducing the number of dimensions without much loss of information.

### 2.2. Self Organizing Map (SOM)

Chattopadhyay *et al.,* (2010) applied the SOM in combination of clustering and projection techniques for part-family and machine-cluster extraction, visualization and interpretation of large high-dimensional machine-part incidence matrix datasets.

The Kohonen SOM is a popular Artificial Neural Network (ANN) based on unsupervised learning having both the properties of vector quantization (Gray, 1984) and vector projection (Kaski, 1997). The use of SOM has been reported primarily in performing the tasks of clustering, pattern recognition and various other optimizations (Kulkarni and Kiang, 1995). The self organization is a process of SOM which configures the output units into a topological representation of the original data, positioning the prototype vectors on a regular low dimensional grid in an ordered fashion making the SOM a powerful visualization tool (Kohonen, 1997). The SOM can project high-dimensional data into low-dimensional space (usually two-dimensional) and meanwhile preserve the original topological relationship. There are few dozen to several thousand neurons connected to adjacent neurons by a neighbourhood relation, dictating the topology of the map. Each neuron i has an associated d-dimensional prototype or codebook vector, $m_i = [m_{i1}, m_{i2},...,m_{id}]$. The dimension d and the input vector are same.

The SOM is trained iteratively with sample vector chosen randomly from the input data set. The Euclidian distance is a typical distance measure to calculate the similarity between sample vector and all prototypes of the map. The Best-Matching Unit (BMU) is that map unit whose incoming connection weights have the greater similarity with the input pattern x and denoted as c (Vesanto, 2002):

$$\|x - m_c\| = \min_i \{\|x - m_i\|\}, \qquad (4)$$

where $\|.\|$ is the distance measure.

After finding the BMU, the prototype vectors (or connection weights) of the SOM are updated. SOM creates a topological mapping by updating not only the BMU's weights, which are adjusted (i.e. moved in the direction of the input pattern by a factor determined by the learning rate), but also adjusting the weights of the adjacent output units in close proximity to the neighborhood of the winner. Thus, not only the BMU get updated, but also the whole neighbourhoods of output neurons are moving closer to the input pattern as specified by the SOM update rule for the weight vector of neuron i in equation 5:

$$m_i(t+1) = m_i(t) + \alpha(t)h_{ci}(r(t))[x(t) - m_i(t)], \qquad (5)$$

Where $\alpha(t)$ is the learning rate, that is a monotonically decreasing function of time t and lie in the range [0, 1], and $h_{ci}(r(t))$ is the neighbourhood kernel around the BMU c, with neighbourhood radius $r(t)$, which typically decreases with time.

## 3. Performance Measure

A measure known as Group Technology Efficiency (GTE) developed by Harhalakis *et al.,* (1990) can be used to evaluate the performance of goodness of the block diagonal form of output matrix from cell formation problem based on sequence of operation. The GTE is defined as the ratio of the difference between the maximum number of inter-cell travels possible and the numbers of inter-cell travels actually required by the system to the maximum number of inter cell travels possible as given in the equation (8). The maximum numbers of inter-cell travels possible in the system is

$$I_p = \sum_{j=1}^{N} (n_o - 1) \qquad (6)$$

The number of inter-cell travels required by the system is

$$I_r = \sum_{j=1}^{N} \sum_{w=1}^{n_o - 1} t_{njw} \qquad (7)$$

Thus the GTE is calculated as

$$GTE = \frac{I_p - I_r}{I_p} \qquad (8)$$

Where, $I_p$ = maximum number of intercell travel possible in the system
$I_r$ = number of intercell travel actually required by the system
n = number of operations (w=1, 2, 3… n)

$$t_{njw} \begin{cases} = 0 \text{ if the operations } w, w+1 \text{ are performed in the same cell} \\ = 1 \text{ otherwise} \end{cases}$$

The GTE is powerful performance measure. The more the value of GTE the better the goodness of cell formation based on operation sequence.

## 4. Experimental Result and Discussion

### 4.1. *PCA extraction and Projection*

PCA has extracted the 16X16 and 14X14 dimension as the parts and machines variable respectively from machine-part incidence matrix based on sequence dataset using correlation matrix. Table 5 and table 6 give summary of eigenvalues and the variances of data from the first five PCs. The eigenvalues illustrated that the first two PCs explain 80% and 85.7% cumulative variance of the parts and machine datasets respectively. The higher percentage of the variance indicated that these PCs are appropriately used to explain the similarity pattern of each part and machine based on their operation sequence information for different parts and machines. Therefore, the original variables (both parts and machines) can be weighted as a linear combination structure into PC1 and PC2 for both parts and machines as follows:

For machines:
$$PC1 = -0.20(m1) + 0.33(m2) - 0.22(m3) + 0.38(m4) - 0.24(m5) + 0.35(m6) - 0.18(m7) +$$
$$0.38(m8) - 0.20(m9) + 0.29(m10) - 0.21(m11) - 0.22(m12) - 0.13(m13) - 0.26(m14) \quad (9)$$
$$PC2 = -0.36(m1) + 0.02(m2) + 0.24(m3) - 0.01(m4) - 0.29(m5) - 0.01(m6) + 0.37(m7)$$
$$- 0.01(m8) - 0.36(m9) - 0.17(m10) + 0.35(m11) + 0.35(m12) - 0.37(m13) - 0.22(m14) \quad (10)$$

The interpretation of the PCs could be made with the help of the absolute value of the coefficient. For example,
in the equation (9) the PC1 interpreted that the machine variables m2, m4, m6, m8, m10 with largest positive coefficient are more similar to form a machine cluster 1.

- in the equation (10) the PC2 interpreted that the machine variables m1, m5, m9, m13, m14 with largest negative coefficient are more similar to form machine cluster 2.
- in the equation (10) the PC2 interpreted that the machine variables m3, m7, m11, m12 with largest positive coefficient are more similar to form another machine cluster 3.

For parts:
$$PC1 = 0.19(p1) - 0.32(p2) + 0.19(p3) - 0.04(p4) + 0.21(p5) - 0.04(p6) - 0.34(p7) + 0.21(p8) -$$
$$0.27(p9) + 0.22(p10) + 0.14(p11) - 0.34(p12) - 0.35(p13) - 0.34(p14) - 0.34(p15) + 0.14(p16) \quad (11)$$

PC2 = 0.34(p1) + 0.14(p2) + 0.34(p3) - 0.34(p4) + 0.20(p5) - 0.34(p6) + 0.09(p7) + 0.20(p8) + 0.02(p9) + 0.32(p10) - 0.39(p11) + 0.07(p12) + 0.08(p13) + 0.09(p14) + 0.12(p15) - 0.39(p16)     (12)

The interpretation of the PCs could be made with the help of the absolute value of the coefficient. For example,
- in the equation (11) the PC1 interpreted that the part variables p2, p7, p9, p12, p13, p14, p15 with largest negative coefficient are more similar to form part family 1.
- in the equation (12) the PC2 interpreted that the part variables p4, p6, p11, p16 with largest negative coefficient are more similar to form part family 2.
- in the equation (12) the PC2 interpreted that the part variables p1, p3, p5, p8, p10 with largest positive coefficient are more similar to form another part family 3.

The loading plot (figure 2a, 2b) provides information about the loadings of the first two principal components. For the Table 1 data pertaining to machine the clustering information of machines (three clusters) are similar as discussed above as evidence from component loading plot in figure 2a. Similarly, for part three visually obvious part families are formed as in the earlier discussion from the component loading plot in figure 2b.

Although the present result revealed that PCA explained most of the cumulative variance of the data but sometimes the component loading plot is difficult to interpret. The component loading plots of a literature based problem in Table 7 are shown in figure 4a and figure 4b. The clustering of machine from figure 4a is not so obvious based on PC projection in 2D plot. The same interpretation for part family identification is also difficult from the component loading plot in figure 4b. This difficulty is primarily due to poor representation on the PCA plane from the complexity of data as reported by Bross *et al.,* (2001). Therefore, an artificial intelligent based unsupervised learning of SOM algorithm is proposed in this present work as an alternative dimension reduction methodology for clustering machine-part cell formation.

### 4.2. *Data Visualisation by SOM*

Unified distance matrix (U-matrix) and Component Plane (CP) are two types of SOM visualization in 2-D hexagonal grid nodes map. There are no explicit rules for choosing the number of nodes in SOM map grid (Hautaniemi *et al.*, 2003) except that the size should allow the easy visualization of the structure of the SOM (Wilppu, 1997). Both of the visualization of data by SOM is discussed below.

Individual component plane maps provide clear visualization of the different input variables. They are displayed in various shades of colors or grey scale color on the SOM component maps. Here each CP represents the machine variable that measured based on the average of operation sequence ordinal values required to be processed by each parts visiting that machine. The component values are de-normalized so that the values shown in the colorbar are in the original ordinal sequence value range. The level of similarity of operation sequence of parts visiting the each machine can be studied from the density of color shades in the hexagonal grid nodes for each CP of SOM map. A darker shade corresponds to larger ordinal values, a grey shade represents a medium ordinal value and a lighter shade represents a low ordinal value of operation sequence. From the observation of CP visualization the possible correlations between the machine variables can also be identified (Table 3). Even the partial correlations can be identified from the inspection of the grayscale representation of the CP. The figure 1 shows that color shades in CP of m1, m5, m9, m13, m14 are almost similar distribution which is reflected from the analysis of correlation of machine (Table 3). In the same way observation of color shades in CPs of m2, m4, m6, m8, m10 revealed their similarity in pattern of color extraction and another pattern of color extraction is visible from the CPs of m3, m7, m11, m12 and both of these pattern can also be supported from the analysis correlation of machines in Table 3. For example, the highest correlation is 1 in m1 and m9 and then 0.90 correlations in m5 and 0.88 correlations in m13 and lowest one is m14 which can be well interpreted by the respective CP visualization.

In figure 1 the top left corner showed a visualization of the U-matrix representing the relative distance measure between the SOM map grid nodes with color extraction. The larger distance between the adjacent nodes represents the darker shades which mean a boundary between clusters. In contrary, regions similar to each other representing a lighter shades means machines are similar indicating a within cluster and thus show a machine cluster. Thus from this U-matrix we can determine three cluster members by locating thick walls between machine-part clusters of figure1 viz.,

machine cluster 1: m2, m4, m6, m8, m10
machine cluster 2: m1, m5, m9, m13, m14
machine cluster 3: m3, m7, m11, m12

Although sometimes it may be seen that SOM map of U-matrix indicating the distance measure is not so reliable to provide a representative machine cluster. Kiang *et al.*,(2004) reported that it may happen in some situation to get a visual cluster is very difficult due to dense data in SOM map. Then the combined visualization of both the U-matrix and CP helps to get both cluster structure and correlations between the variables (figure 1).

Also the labels of parts from sequence based machine part matrix of each unit are shown on an empty grid in the figure 1 at the bottom right position of the panel. Based on the position of the extracted parts in the SOM map grid from this Labels and the visualization combining the U-matrix and CP the parts which are required to be processed by the corresponding machines cluster can be easily identified from the figure 1. Thus the machine cluster 1 corresponds to the part family 1 with the parts p2, p7, p9, p12, p13, p14, p15 in cell 1; machine cluster 2 corresponds to part family 2 with the parts p4, p6, p11, p16 in cell 2; and machine cluster 3 corresponds to part family 3 with the parts p1, p3, p5, p8, p10 in cell 3 are clearly visible from the combined visualization of U-matrix, CP and Labels of figure 1.

The most impressive and fascinating of visual clustering is the figure 3 which is composed of three figures: Color code, PC projection and Labels (from left side) which are outputs of PC projection of SOM. The color code represents the extraction of color based on similarity of parts which are topology preserved. The color code of SOM shows clearly three types of color extraction red, yellow and green with numbers of hit in each node. The hits represent number of part variable extracted based on similar color extraction. The labels diagram of figure 3 in combination with the other two figure helps to identify the formation of part family. Thus three part family visually identified from the clustering information based SOM based PC color extraction. For example green corresponds to part family 1(p2, p7, p9, p12, p13, p14, p15); red corresponds to part family 2 (p4, p6, p11, p16) and yellow corresponds to part family 3 (parts p1, p3, p5, p8, p10). The above clustering of part family can also be validated from Table 4 and the correlation of part family shows highest (p1 and p3) 1.0; second highest (p2 and p15) 0.98; third highest (p1 and p10; p2 and p7; p2 and p14) 0.97 and lowest (p1 and p5; p2 and p9) 0.64.

In this way the block diagonal output of the sequence based cell formation (Table 2) is derived through the visual clustering of U-matrix, component map and PC projection of SOM.

The goodness of cell formation is calculated from the block diagonal form of table 2 using the performance measure, group technology efficiency (GTE). The GTE comes out to be 81.16%.

The SOM clustering approach has also been extended to implement using another sequence based problem from literature (Table 7). The same way using the figure 5 and figure 6 visual clustering of the cell formation is performed to obtain the output block diagonal matrix of Table 8. The U-matrix, CP in figure 5 and PC projection of SOM in figure 6 clearly shows the 4 distinct clusters. The GTE is calculated to be 93.87 from the Table 8.

### 4.3. *Comparison of PCA and SOM methods*

The use of PCA and SOM methods is compared based on data extraction procedure where the applied sequence based machine-part incidence matrix dataset illustrated that both methods suitably extracted the high dimensional data onto a low dimensional representation.

- As the first two PCs provided a high percentage variance for which PCA was an excellent dimensional reduction technique through it's in depth explanation of variance of machine-part dataset based on ordinal data (operation sequence). On the other way the SOM showed the variance of input ordinal data through the visualization of the clustering using the relative distances between the SOM nodes based on the colour extraction on the U-matrix and CP.
- The PCA may not be a suitable method in clustering large dimensional sequence based machine-part incidence matrix in easy and user friendly way because the visual extraction of clustering information is become very difficult from 2-D PC projection plot. In contrast, the SOM can be implemented in a flexible way to cluster the machine-part cell formation even from the complex and large dimensional cell formation problem. The SOM map is very useful to obtain visual clustering information of machine-part cell formation with the help of color extraction of U-matrix, CP, and SOM projection. The SOM projection can efficiently visually cluster large dimensional sequence based machine-part incidence matrix.

Brosse *et al.,* (2001) reported that SOM is provides more reliable data presentation when complex data sets are used. In another study it is reported that SOM is a good visual clustering tool in large dimensional problem (Laitinen *et al.,* 2002; Chattopadhyay *et al.,* 2010). Thus in the context of the present research the SOM can very efficiently visually cluster the large dimensional problem even based on ordinal data. In the present paper a new solution approach to the cell formation problem of operation sequence based machine-part incidence matrix has been presented after the cluster visualization and analysis of U-matrix, component plane and PC projection of SOM.

### 4.4. *Benefits from the present work*

PCA is used to remove multicollinearity from a given set of data consisting of different variables to be used as independent variable for modeling or predicting a dependent variable (Sousa *et al.,* 2007 and references therein). Application of PCA to the preprocessing of data for generating an ANN model is well documented in literature (Abdul-Wahab *et al*, 2005). In such problems the factor loadings can be used to remove the multicollinearity and consequently the complexity of a given data set to be exposed to the ANN framework. In the cellular manufacturing the application of PCA has opened the new avenues to make simultaneously machine groups and part families while maximizing correlation between elements (Hachicha *et al.,* 2006, 2008; Arvindh and Irani, 1991). In the present work we have discerned that the use of PCA and SOM approaches suitably extracted the high dimensional data onto a low dimensional representation, which is consistent with the earlier investigations on PCA with respect to its dimension reduction capacity. Moreover, it is quite appealing that the conventional PCA is having similar potential to that of PCA in dealing with the present problem.

### 5. Conclusions

SOM and PCA have been implemented to visualize and cluster machine-part cell formation based on ordinal sequence data. Although the numerically illustrated result revealed that PCA has explained most of the cumulative variance of data but in reality when the very large dimensional machine-part cell formation problem based on sequence is available then to obtain the clustering structure from PCA projection has become very difficult. Most importantly, the CP and SOM projection with color extraction and labeling is effective tool to visually cluster machine-part cell formation. In the visual clustering of ordinal data, the use of U-matrix alone can not be efficient to get the cluster structure but with color extraction, hit map, labeling via the SOM node map it becomes a powerful simultaneous clustering visualization methodology of machines and parts. Thus identification of part family to be processed by machine cluster can be easily interpreted from the SOM map visualization. Sometimes coloring is not so clear to designate the cluster border for which visual clustering of SOM is a challenge.

### Acknowledgement

The authors of the present paper acknowledge the inspiration and motivation shown by Dr. Surajit Chattopadhyay of Pailan College of Management & Technology an affiliated college under West Bengal University of Technology during the entire period of this work. Sincere thanks are due to the anonymous reviewer for giving constructive comments.

**References**

Abdul-Wahab, S.A., Bakheit, C.S., Al-Alawi, S.M. [2005] Principal component and multiple regression analysis in modelling of ground-level ozone and factors affecting its concentrations, Environmental Modelling & Software, 20(10), 1263-1271

Ampazis, N., Minis, I. [2004] Design of cellular manufacturing systems using Latent Semantic Indexing and Self Organizing Maps, Computational Management Science, 1(3-4), 275-292

Andrés, C., Lozano, S. [2006] A particle swarm optimization algorithm for part–machine grouping, Robotics and Computer-Integrated Manufacturing, 22(5-6), 468-474

Annas, S., Kanai, T., and Koyama, S. [2007] Principal Component Analysis and Self-Organizing Map for visualizing and classifying fire risks in forest regions, Agricultural Information Research. 16(2), 44-51

Arvindh, B., Irani, S. A. [1991] Principal components analysis for evaluating the feasibility of cellular manufacturing without initial-part matrix clustering. International Journal of Production Research. 32, 1909-1938.

Ateme-Nguema, B., Dao, T-M. [2009] Quantized Hopfield networks and tabu search for manufacturing cell formation problems, International Journal of Production Economics, 121(1), 88-98

Atsma, W.J., Hodgson, A.J. [1999] Inferring motor plan complexity using a modified principal component analysis, Engineering in Medicine and Biology, 21st Annual Conference 1: 533

Barbieri, P., Adami, G., Piselli, S., Gemiti, F., Reisenhofer, E. [2002] A three-way principal factor analysis for assessing the time variability of freshwaters related to a municipal water supply, Chemometrics and Intelligent Laboratory Systems, 62(1), 89-100

Baykasoglu, A. and Gindy, N.N.Z. [2000] MOCACEF: multiple objective capability based approach to form part machine groups for cellular manufacturing application, Int. J. Prod. Res., 38(5), 1133–1161

Brosse, S., Giraudel, J.L., and Lek, S. [2001] Utilisations of non-supervised neural networks and principal component analysis to study fish assemblages, Ecological modeling, 146, 159-166

Burbidge, J.L. [1963] Production flow Analysis, Production Engineer, 742-752.

Burbidge, J.L. [1971] Production flow Analysis, Production Engineer, 50(4/5), 139-152.

Bu, W., Liu, Z., Tan, J. [2009] Industrial robot layout based on operation sequence optimization, International Journal of Production Research, 47(15), 4125 – 4145

Chan, H.M. and Milner, D. A. [1982] Direct Clustering Algorithm For Group Formation in Cellular Manufacture, Journal Of Manufacturing Systems, 1, 65- 75

Çakar, T., Cil, I. [2004] Artificial neural networks for design of manufacturing systems and selection of priority rules, International Journal of Computer Integrated Manufacturing, 17(3), 195 – 211

Chattopadhyay, M., Chattopadhyay, S., Dan, K. P. [2010] Machine-Part cell formation through visual decipherable clustering of Self Organizing Map, 52(9-12), 1019-1030, DOI: 10.1007/s00170-010-2802-4

Chu, C-H. [1993] Manufacturing cell formation by competitive learning, International Journal of Production Research, 31(4), 829 - 843

Choobineh, F. [1988] A framework for the design of cellular manufacturing systems, International Journal of Production Research, 26, 1161-1172.

Defersha, F. M., Chen, M. [2006] A comprehensive mathematical model for the design of cellular manufacturing systems, International Journal of Production Economics, 103(2), 767-783

Dimopoulos, C. [2006] Multi-objective optimization of manufacturing cell design International Journal of Production Research, 44(22), 4855 - 4875

George, A. P., Rajendran, C., Ghosh, S. [2003] An analytical-iterative clustering algorithm for cell formation in cellular manufacturing systems with ordinal-level and ratio-level data, The International Journal of Advanced Manufacturing Technology, 22(1-2), 125-133

Gray, R. M. [1984] Readings in speech recognition, edited by Alex Waibel, Kai-Fu Lee, Morgan Kaufmann Publishers Inc., San Francisco, CA, USA, 75 - 100

Gunasekharan, A., Goyal, S.K., Virtanen, I., Yli-Olli, P. [1994]. An investigation into the application of group technolog in advanced manufacturing systems, Int. J. Computer Integrated Manufacturing, 7(4), 215-228

Hachicha, W., Masmoudi, F., Haddar, M. [2006] A correlation analysis approach of cell formation in cellular manufacturing system with incorporated production data, International Journal of Manufacturing Research, 1(3), 332 – 353

Hachicha, W., Masmoudi, F., Haddar, M. [2008] Formation of machine groups and part families in cellular manufacturing systems using a correlation analysis approach, International Journal of Advanced Manufacturing Technology, 36(11-12), 1157-1169

Harhalakis, G., Nagi, R., Proth, J.M. [1990] An efficient heuristic in manufacturing cell formation for group technology applications, International Journal of Production Research, 28(1),185–198.

Hautaniemi, S., Yli-Harja, O., Astola, J., Kauraniemi, P., Kallioniemi, A., Wolf, M., Ruiz, J., Mousses, S., Kallioniemi, O-P. [2003] Analysis and Visualization of Gene Expression Microarray Data in Human Cancer Using Self-Organizing Maps, Machine Learning, 52(1-2), 45-66(22)

Haykin, S. [2008] Neural Networks a comprehensive foundation, 2nd edition, Pearson Education, India, Delhi.

Horenko, I., Dittmer, E., Schütte, C. [2006] Reduced Stochastic Models for Complex Molecular Systems, Computing and Visualization in Science, 9(2), 89-102

Hotelling, H. [1933] Analysis of a Complex of Statistical Variables into Principal Components, The Journal of Educational Psychology, 498-520

Inho, J., Jongtae, R. [1997] Generalized machine cell formation considering material flow and plant layout using modified self-organizing feature maps, Computers & Industrial Engineering, 33(3-4), 457-460

Jayaswal, S., Adil, G. K. [2004] Efficient algorithm for cell formation with sequence data, machine replications and alternative process routings, International Journal of Production Research, 42(12), 2419 - 2433

Johnson, R. A. and Wichren, D.W. [1998] Applied Multivariate Statistical Analysis, International Edition, 4, United States of America: Prentice-Hall, ISBN:0-130-41146-9

Jolliffe, I. T. [1986] Principal component analysis, New York: Springer, ISBN: 0-387-96269-7.

Kang, S-L. and Wemmerlöv, U. [1993] A work load-oriented heuristic methodology for manufacturing cell formation allowing reallocation of operations, European Journal of Operational Research, 69(3), 292-311

Kaski, S. [1997] Data exploration using self-organizing maps, Acta Polytecna Scandinavica, Mathematics, Computing and Management in Engineering Series No. 82, Espoo 1997, 57 pp. Published by Finish Academy of technology, ISBN 952-5148-13-0, ISSN 1238-9803, UDC 681.327.12.159.95.519.2

Kiang, M.Y., Kulkarni, U.R. and Tam, K.Y. [1995] Self-organising map network as an interactive clustering tool – an application to group technology, Decision Support Systems, 15, 351-374

Kiang, M.Y., Kumar, A. A. [2004] comparative analysis of an extended SOM network and K-means analysis, Journal International Journal of Knowledge-Based and Intelligent Engineering Systems, 8(1/2004), 9-15

Kohonen, T. [1997] Self-Organizing Maps, Springer, Berlin.

Koua, E.L. [2003] Using self-organizing maps for information Visualization and knowledge discovery in complex geospatial datasets,Proceedings of the 21st International Cartographic Conference (ICC) Durban, South Africa, 1694-1702

Kulkarni, U. R. and Kiang, Y. M. [1995] Dynamic grouping of parts in flexible manufacturing systems — a self-organizing neural networks approach, European Journal of Operational Research, 84(1), 192-212

Kwan,,C., Xu, R. and Hayness, L. [2001] A new data clustering and its applications. In proceeding of SPIE-The International Society for Optical Engineering, 4384, 1-5

Laitinen, N., Rantanen, J., Laine, S., Osmo, A., Räsänen, E., Sari, A. and Jouko, Y. [2002] Visualization of particle size and shape distributions using self-organizing maps, Chemometrics and Intelligent Laboratory Systems, 62(1), 47-60

Lee-Post, A. [2000] Part family identification using a simple genetic algorithm. Int. J. Prod. Res., 38(4), 793–810.

Li, X., Baki, M.F., Aneja, Y.P. [2010] An ant colony optimization metaheuristic for machine–part cell formation problems, Computers & Operations Research, 37[12], 2071-2081

Lin, S.W., Ying, K-C., and Lee, Z-J. [2010] Part-machine cell formation in group technology using a simulated annealing-based meta-heuristic, International Journal of Production Research, 48, 3579 – 3591

Mahdavi, I., and Mahadevan, B. [2008] CLASS: An algorithm for cellular manufacturing system and layout design using sequence data, Robotics and Computer-Integrated Manufacturing archive, 24[3], 488-497

Mahdavi, I., Aalaei, A., Paydar, M. M., and Solimanpur, M. [2010] Designing a mathematical model for dynamic cellular manufacturing systems considering production planning and worker assignment, Computers & Mathematics with Applications, 60(4), 1014-1025 doi:10.1016/j.camwa.2010.03.044

Mukhopadhyay, S. K., Babu, K. R. and Sai, K.V.V. [2009] A note on 'Modified Hamiltonian chain: a graph theoretic approach to group technology', International Journal of Production Research, 47(1), 289 – 298

Nagi, R., Harhalakis, G., Proth, J-M. [1990] Multiple routeings and capacity considerations in group technology applications, International Journal of Production Research, 28(12), 1243 – 1257

Nair, G.J., and Narendran,T.T. [1998] CASE: A Clustering Algorithm for Cell Formation with sequence data. International journal of Production Rsearch, 36(1), 157-179

Ohta, H. and Nakamura, M. [2002] Cell formation with reduction in setup timesComputers & Industrial Engineering, 42[2-4], 317-327

Oliveira, S., Ribeiro, J.F.F., Seok, S.C. [2008] A comparative study of similarity measures for manufacturing cell formation, Journal of Manufacturing Systems, 27[1], 19-25

Pandian, S. R., Mahapatra, S.S. [2008] Cell formation with ordinal-level data using ART1-based neural networks, Int. J. Services and Operations Management, 4(5)

Pandian, S. R., Mahapatra, S.S. [2009] Manufacturing cell formation with production data using neural networks, Computers & Industrial Engineering, 56[4], 1340-1347

Papaioannou, G., Wilson, J.M. [2010] The evolution of cell formation problem methodologies based on recent studies (1997–2008): Review and directions for future research, European Journal of Operational Research, 206[3], 509-52

Pearson, K. [1901] On lines and planes of closest fit to systems of points in space. Philosophical Magazine, 2, 559-572

Pollalis, AY., Mavrommatis, G. [2009] Using similarity measures for collaborating groups formation: A model for distance learning environments,European Journal of Operational Research, 193[2], 626-636

Preisendorfer, R. [1988] Principal component analysis in meteorology and oceanography, Elsevier, Amsterdam, NL, 425

Rogers, D. F., Kulkarni, S, S. [2005] Optimal bivariate clustering and a genetic algorithm with an application in cellular manufacturing, European Journal of Operational Research, 160(2), 423-444

Rothenberger, M.A., Dooley, K. J., Kulkarni, U.R., Nada, N. [2003] Strategies for Software Reuse: A Principal Component Analysis of Reuse Practices, IEEE Transactions on Software Engineering, 29(9), 825-837.

Selvam, R. P., and Balasubramanian, K.N. [1985] Algorithmic grouping of operation sequences, Engineering Costs and Production Economics, 9(1-3), 125-134

Soleymanpour, M., Vrat, P., and Shankar, R. [2002] A transiently chaotic neural network approach to the design of cellular manufacturing, International Journal of Production Research, 40[10], 2225 – 2244

Sousa, S.I.V., Martins, F.G., Alvim-Ferraz, M.C.M., Pereira, M.C. [2007] Multiple linear regression and artificial neural networks based on principal components to predict ozone concentrations, Environmental Modelling & Software, 22(1), 97-103

Tipping, M.E., and Bishop, C.M., [1999] Mixtures of probabilistic principal component analysers, Neural Computation,11, 443-482

Tuncer, Y., Tanik, M. M., Alison, D.B. [2008] An overview of statistical decomposition techniques applied to complex systems, Computational Statistics & Data Analysis, 52(5), 2292-2310.

Ünler, A., Güngör, Z. [2009] Applying K-harmonic means clustering to the part-machine classification problem, Expert Systems with Applications, 36(2), Part 1, 1179-1194

Vakharia, A. J., Wemmerlov, U. [1990] Designing a Cellular Manufacturing System: A Materials Flow Approach Based on Operation Sequences, IIE Transactions, 22(1), 84 – 97

Venkataramanaiah, S. [2008] Scheduling in cellular manufacturing systems: an heuristic approach, International Journal of Production Research, 46(2), 429 – 449

Venkumar, P., Haq, A. N. [2006] Complete and fractional cell formation using Kohonen self-organizing map networks in a cellular manufacturing system, International Journal of Production Research, 44(20), 4257 – 4271

Venugopal, V., and Narendran, T. T. [1994] Machine-cell formation through neural network models, International Journal of Production Research, 32(9), 2105 – 2116

Verleysen, M. [2003] Learning high-dimensional data, Limitations and Future Trends in Neural Computation, S. Ablameyko et al. (Eds.), IOS Press, 141-162

Vesanto, J. [2002] Data Exploration Process Based on the Self-Organizing Map, PhD thesis, Helsinki University of Technology, Espoo, Finland.

Wall, M.E., Rechtsteiner, A., Rocha, L.M. [2003] A Practical Approach to Microarray Data Analysis, Kluwer: Norwell, MA, 91-109,

Wilppu, E. [1997] The Visualisation Capability of Self-Organizing Maps to Detect Deviations in Distribution Control, Technical Report: TUCS-TR-153, Turku Centre for Computer Science, ISBN 952-12-0127-4

Won, Y., and Currie, K. R. [2007] Fuzzy ART/RRR-RSS: a two-phase neural network algorithm for part-machine grouping in cellular manufacturing, International Journal of Production Research, 45[9], 2073–2104

Yeung, K. Y., Ruzzo, W. L. [2001] Principal component analysis for clustering gene expression data, Bioinformatics, 17[9], 763-774

Yin, Y., Yasuda, K. [2002] Manufacturing cells' design in consideration of various production factors, International Journal of Production Research, 40[4], 885 – 906

Yin, Y., and Yasuda, K. [2006] Similarity coefficient methods applied to the cell formation problem: A taxonomy and review, International Journal of Production Economics, 101(2),329-352

Zhao, C., and Wu, Z. [2000] A genetic algorithm for manufacturing cell formation with multiple routes and multiple objectives, Int. J. Prod. Res, 38(2), 385–395

Appendix

Table 1. Machine-part incidence matrix with sequence data (artificially generated)

|     | m1 | m2 | m3 | m4 | m5 | m6 | m7 | m8 | m9 | m10 | m11 | m12 | m13 | m14 |
|-----|----|----|----|----|----|----|----|----|----|-----|-----|-----|-----|-----|
| p1  | 0  | 1  | 3  | 0  | 0  | 0  | 4  | 0  | 0  | 0   | 2   | 5   | 0   | 0   |
| p2  | 0  | 6  | 0  | 3  | 0  | 1  | 2  | 4  | 0  | 5   | 0   | 0   | 0   | 0   |
| p3  | 0  | 1  | 3  | 0  | 0  | 0  | 4  | 0  | 0  | 0   | 2   | 5   | 0   | 0   |
| p4  | 5  | 0  | 0  | 0  | 1  | 0  | 0  | 0  | 4  | 6   | 0   | 0   | 2   | 3   |
| p5  | 0  | 0  | 0  | 0  | 0  | 0  | 3  | 0  | 0  | 0   | 1   | 5   | 0   | 4   |
| p6  | 5  | 0  | 0  | 0  | 1  | 0  | 0  | 0  | 4  | 6   | 0   | 0   | 2   | 3   |
| p7  | 0  | 5  | 0  | 2  | 0  | 1  | 0  | 3  | 0  | 4   | 0   | 0   | 0   | 0   |
| p8  | 0  | 0  | 0  | 0  | 0  | 0  | 3  | 0  | 0  | 0   | 1   | 5   | 0   | 4   |
| p9  | 0  | 0  | 0  | 3  | 0  | 2  | 0  | 4  | 0  | 5   | 0   | 0   | 0   | 0   |
| p10 | 0  | 0  | 3  | 0  | 1  | 0  | 4  | 0  | 0  | 0   | 2   | 5   | 0   | 0   |
| p11 | 6  | 0  | 1  | 0  | 2  | 0  | 0  | 0  | 5  | 0   | 0   | 0   | 3   | 4   |
| p12 | 0  | 6  | 0  | 3  | 0  | 1  | 0  | 4  | 0  | 5   | 0   | 0   | 2   | 0   |
| p13 | 0  | 6  | 0  | 3  | 0  | 2  | 0  | 4  | 0  | 5   | 0   | 0   | 1   | 0   |
| p14 | 0  | 6  | 0  | 3  | 0  | 1  | 0  | 4  | 0  | 5   | 0   | 0   | 0   | 0   |
| p15 | 0  | 6  | 0  | 3  | 0  | 2  | 1  | 4  | 0  | 5   | 0   | 0   | 0   | 0   |
| p16 | 6  | 0  | 1  | 0  | 2  | 0  | 0  | 0  | 5  | 0   | 0   | 0   | 3   | 4   |

Table 2. Final output block diagonal form of table 1 data

|     | m11 | m3 | m7 | m12 | m6 | m4 | m8 | m10 | m2 | m5 | m13 | m14 | m9 | m1 |
|-----|-----|----|----|-----|----|----|----|----|----|----|-----|-----|----|----|
| p1  | 2   | 3  | 4  | 5   |    |    |    |     | 1  |    |     |     |    |    |
| p10 | 2   | 3  | 4  | 5   |    |    |    |     |    | 1  |     |     |    |    |
| p3  | 2   | 3  | 4  | 5   |    |    |    |     | 1  |    |     |     |    |    |
| p8  | 1   |    | 3  | 5   |    |    |    |     |    |    |     | 4   |    |    |
| p5  | 1   |    | 3  | 5   |    |    |    |     |    |    |     | 4   |    |    |
| p13 |     |    |    |     | 2  | 3  | 4  | 5   | 6  |    | 1   |     |    |    |
| p15 |     |    | 1  |     | 2  | 3  | 4  | 5   | 6  |    |     |     |    |    |
| p9  |     |    |    |     | 2  | 3  | 4  | 5   |    |    |     |     |    |    |
| p2  |     |    | 2  |     | 1  | 3  | 4  | 5   | 6  |    |     |     |    |    |
| p12 |     |    |    |     | 1  | 3  | 4  | 5   | 6  |    | 2   |     |    |    |
| p7  |     |    |    |     | 1  | 2  | 3  | 4   | 5  |    |     |     |    |    |
| p14 |     |    |    |     | 1  | 3  | 4  | 5   | 6  |    |     |     |    |    |
| p4  |     |    |    |     |    |    |    | 6   |    | 1  | 2   | 3   | 4  | 5  |
| p16 |     | 1  |    |     |    |    |    |     |    | 2  | 3   | 4   | 5  | 6  |
| p11 |     | 1  |    |     |    |    |    |     |    | 2  | 3   | 4   | 5  | 6  |
| p6  |     |    |    |     |    |    |    | 6   |    | 1  | 2   | 3   | 4  | 5  |

Table 3. Correlation matrix of machine variables

|     | m1    | m2    | m3    | m4    | m5    | m6    | m7    | m8    | m9    | m10   | m11   | m12   | m13  | m14  |
|-----|-------|-------|-------|-------|-------|-------|-------|-------|-------|-------|-------|-------|------|------|
| m1  | 1.00  |       |       |       |       |       |       |       |       |       |       |       |      |      |
| m2  | -0.48 | 1.00  |       |       |       |       |       |       |       |       |       |       |      |      |
| m3  | -0.07 | -0.38 | 1.00  |       |       |       |       |       |       |       |       |       |      |      |
| m4  | -0.50 | 0.85  | -0.52 | 1.00  |       |       |       |       |       |       |       |       |      |      |
| m5  | 0.90  | -0.52 | 0.17  | -0.54 | 1.00  |       |       |       |       |       |       |       |      |      |
| m6  | -0.46 | 0.69  | -0.48 | 0.92  | -0.50 | 1.00  |       |       |       |       |       |       |      |      |
| m7  | -0.46 | -0.31 | 0.71  | -0.46 | -0.28 | -0.44 | 1.00  |       |       |       |       |       |      |      |
| m8  | -0.50 | 0.85  | -0.52 | 1.00  | -0.54 | 0.92  | -0.46 | 1.00  |       |       |       |       |      |      |
| m9  | 1.00  | -0.48 | -0.07 | -0.50 | 0.91  | -0.46 | -0.46 | -0.50 | 1.00  |       |       |       |      |      |
| m10 | -0.03 | 0.56  | -0.66 | 0.69  | -0.28 | 0.63  | -0.67 | 0.69  | -0.05 | 1.00  |       |       |      |      |
| m11 | -0.36 | -0.42 | 0.85  | -0.55 | -0.17 | -0.51 | 0.94  | -0.55 | -0.36 | -0.71 | 1.00  |       |      |      |
| m12 | -0.39 | -0.47 | 0.65  | -0.59 | -0.23 | -0.54 | 0.94  | -0.59 | -0.39 | -0.75 | 0.94  | 1.00  |      |      |
| m13 | 0.88  | -0.24 | -0.14 | -0.28 | 0.81  | -0.29 | -0.57 | -0.29 | 0.88  | 0.03  | -0.45 | -0.48 | 1.00 |      |
| m14 | 0.69  | -0.64 | -0.21 | -0.67 | 0.61  | -0.61 | -0.10 | -0.67 | 0.69  | -0.37 | -0.13 | 0.08  | 0.56 | 1.00 |

Table 4. Correlation matrix of part variables

|     | p1    | p2    | p3    | p4    | p5    | p6    | p7    | p8    | p9    | p10   | p11   | p12   | p13   | p14   | p15   | p16  |
|-----|-------|-------|-------|-------|-------|-------|-------|-------|-------|-------|-------|-------|-------|-------|-------|------|
| p1  | 1.00  |       |       |       |       |       |       |       |       |       |       |       |       |       |       |      |
| p2  | -0.18 | 1.00  |       |       |       |       |       |       |       |       |       |       |       |       |       |      |
| p3  | 1.00  | -0.18 | 1.00  |       |       |       |       |       |       |       |       |       |       |       |       |      |
| p4  | -0.47 | -0.03 | -0.47 | 1.00  |       |       |       |       |       |       |       |       |       |       |       |      |
| p5  | 0.64  | -0.28 | 0.64  | -0.16 | 1.00  |       |       |       |       |       |       |       |       |       |       |      |
| p6  | -0.47 | -0.03 | -0.47 | 1.00  | -0.16 | 1.00  |       |       |       |       |       |       |       |       |       |      |
| p7  | -0.28 | 0.97  | -0.28 | 0.03  | -0.36 | 0.03  | 1.00  |       |       |       |       |       |       |       |       |      |
| p8  | 0.64  | -0.28 | 0.64  | -0.16 | 1.00  | -0.16 | -0.36 | 1.00  |       |       |       |       |       |       |       |      |
| p9  | -0.38 | 0.64  | -0.38 | 0.18  | -0.33 | 0.18  | 0.63  | -0.33 | 1.00  |       |       |       |       |       |       |      |
| p10 | 0.97  | -0.30 | 0.97  | -0.45 | 0.64  | -0.45 | -0.41 | 0.64  | -0.38 | 1.00  |       |       |       |       |       |      |
| p11 | -0.41 | -0.53 | -0.41 | 0.65  | -0.07 | 0.65  | -0.47 | -0.07 | -0.43 | -0.36 | 1.00  |       |       |       |       |      |
| p12 | -0.34 | 0.93  | -0.34 | 0.04  | -0.41 | 0.04  | 0.97  | -0.41 | 0.64  | -0.47 | -0.43 | 1.00  |       |       |       |      |
| p13 | -0.34 | 0.95  | -0.34 | 0.01  | -0.41 | 0.01  | 0.99  | -0.41 | 0.68  | -0.47 | -0.48 | 0.98  | 1.00  |       |       |      |
| p14 | -0.29 | 0.97  | -0.29 | 0.02  | -0.36 | 0.02  | 1.00  | -0.36 | 0.67  | -0.42 | -0.47 | 0.97  | 0.99  | 1.00  |       |      |
| p15 | -0.26 | 0.98  | -0.26 | -0.03 | -0.34 | -0.03 | 0.99  | -0.34 | 0.68  | -0.38 | -0.53 | 0.95  | 0.98  | 0.99  | 1.00  |      |
| p16 | -0.41 | -0.53 | -0.41 | 0.65  | -0.07 | 0.65  | -0.47 | -0.07 | -0.43 | -0.36 | 1.00  | -0.43 | -0.48 | -0.47 | -0.53 | 1.00 |

Table 5. Eigen value extracted of first 5 components for parts

| Component | Eigenvalue | Proportion | Cumulative |
|---|---|---|---|
| 1 | 7.79 | 0.49 | 0.49 |
| 2 | 5.00 | 0.31 | 0.80 |
| 3 | 1.64 | 0.10 | 0.90 |
| 4 | 0.69 | 0.04 | 0.95 |
| 5 | 0.65 | 0.04 | 0.99 |

Extraction method by PCA

Table 6. Eigen value extracted of first 5 components for machines

| Component | Eigenvalue | Proportion | Cumulative |
|---|---|---|---|
| 1 | 6.74 | 0.48 | 0.48 |
| 2 | 5.26 | 0.38 | 0.86 |
| 3 | 0.92 | 0.07 | 0.92 |
| 4 | 0.41 | 0.03 | 0.95 |
| 5 | 0.33 | 0.02 | 0.98 |

Extraction method by PCA

Table7. Machine-part incidence matrix with sequence data from Nagi et al.,1990

|  | m1 | m2 | m3 | m4 | m5 | m6 | m7 | m8 | m9 | m10 | m11 | m12 | m13 | m14 | m15 | m16 | m17 | m18 | m19 | m20 |
|---|---|---|---|---|---|---|---|---|---|---|---|---|---|---|---|---|---|---|---|---|
| p1  | 0 | 0 | 0 | 0 | 0 | 0 | 3 | 0 | 2 | 0 | 0 | 1 | 0 | 0 | 0 | 0 | 0 | 0 | 0 | 0 |
| p2  | 3 | 0 | 0 | 0 | 0 | 0 | 1 | 0 | 0 | 0 | 0 | 2 | 0 | 0 | 0 | 0 | 0 | 0 | 0 | 0 |
| p3  | 1 | 0 | 0 | 0 | 0 | 0 | 4 | 0 | 2 | 0 | 0 | 3 | 0 | 0 | 0 | 0 | 0 | 0 | 0 | 0 |
| p4  | 1 | 0 | 0 | 0 | 0 | 0 | 3 | 0 | 0 | 0 | 0 | 2 | 0 | 0 | 0 | 0 | 0 | 0 | 0 | 0 |
| p5  | 2 | 0 | 0 | 0 | 0 | 0 | 0 | 0 | 1 | 0 | 0 | 3 | 0 | 0 | 0 | 0 | 0 | 5 | 6 | 4 |
| p6  | 0 | 3 | 0 | 0 | 2 | 1 | 0 | 0 | 0 | 0 | 0 | 0 | 0 | 0 | 0 | 0 | 0 | 0 | 0 | 0 |
| p7  | 0 | 0 | 0 | 0 | 1 | 4 | 0 | 0 | 0 | 0 | 0 | 0 | 0 | 0 | 0 | 2 | 0 | 0 | 3 | 0 |
| p8  | 0 | 3 | 0 | 0 | 0 | 2 | 0 | 0 | 0 | 0 | 0 | 0 | 0 | 0 | 0 | 1 | 0 | 0 | 0 | 0 |
| p9  | 0 | 2 | 0 | 0 | 0 | 4 | 0 | 0 | 0 | 0 | 0 | 0 | 0 | 0 | 0 | 1 | 0 | 0 | 3 | 0 |
| p10 | 0 | 1 | 0 | 0 | 3 | 4 | 0 | 0 | 0 | 0 | 0 | 0 | 0 | 0 | 0 | 2 | 0 | 0 | 0 | 0 |
| p11 | 0 | 0 | 2 | 0 | 0 | 0 | 0 | 1 | 0 | 0 | 3 | 0 | 0 | 0 | 0 | 0 | 0 | 4 | 0 | 0 |
| p12 | 0 | 0 | 1 | 0 | 0 | 0 | 0 | 2 | 0 | 0 | 0 | 0 | 0 | 0 | 0 | 0 | 0 | 3 | 0 | 0 |
| p13 | 0 | 0 | 3 | 0 | 0 | 0 | 0 | 2 | 0 | 0 | 1 | 0 | 0 | 0 | 0 | 0 | 0 | 4 | 0 | 0 |
| p14 | 0 | 0 | 0 | 0 | 0 | 0 | 0 | 0 | 0 | 1 | 0 | 0 | 0 | 3 | 0 | 0 | 4 | 0 | 0 | 2 |
| p15 | 0 | 0 | 0 | 0 | 0 | 0 | 0 | 0 | 0 | 3 | 0 | 0 | 0 | 0 | 0 | 0 | 2 | 0 | 0 | 1 |
| p16 | 0 | 0 | 0 | 0 | 0 | 0 | 0 | 0 | 0 | 2 | 0 | 0 | 0 | 3 | 0 | 0 | 0 | 0 | 0 | 1 |
| p17 | 0 | 0 | 0 | 0 | 0 | 0 | 0 | 0 | 0 | 3 | 0 | 0 | 0 | 2 | 0 | 0 | 1 | 0 | 0 | 0 |
| p18 | 0 | 0 | 0 | 2 | 0 | 0 | 0 | 0 | 0 | 0 | 0 | 1 | 0 | 3 | 0 | 0 | 0 | 0 | 0 | 0 |
| p19 | 0 | 0 | 0 | 1 | 0 | 0 | 0 | 0 | 0 | 0 | 0 | 0 | 2 | 3 | 0 | 0 | 0 | 0 | 0 | 0 |
| p20 | 0 | 0 | 0 | 2 | 0 | 0 | 0 | 0 | 0 | 0 | 0 | 0 | 0 | 0 | 1 | 0 | 0 | 0 | 0 | 0 |

Table 8. Block diagonal out put of the sequence based machine-part incidence matrix from Nagi et al., 1990

| | m2 | m6 | m5 | m16 | m19 | m1 | m7 | m9 | m12 | m3 | m8 | m11 | m18 | m20 | m4 | m13 | m15 | m10 | m14 | m17 |
|---|---|---|---|---|---|---|---|---|---|---|---|---|---|---|---|---|---|---|---|---|
| p6  | 3 | 1 | 2 |   |   |   |   |   |   |   |   |   |   |   |   |   |   |   |   |   |
| p9  | 2 | 4 |   | 1 | 3 |   |   |   |   |   |   |   |   |   |   |   |   |   |   |   |
| p7  |   | 4 | 1 | 2 | 3 |   |   |   |   |   |   |   |   |   |   |   |   |   |   |   |
| p8  | 3 | 2 |   | 1 |   |   |   |   |   |   |   |   |   |   |   |   |   |   |   |   |
| p10 | 1 | 4 | 3 | 2 |   |   |   |   |   |   |   |   |   |   |   |   |   |   |   |   |
| p1  |   |   |   |   |   |   | 3 | 2 | 1 |   |   |   |   |   |   |   |   |   |   |   |
| p3  |   |   |   |   |   | 1 | 4 | 2 | 3 |   |   |   |   |   |   |   |   |   |   |   |
| p4  |   |   |   |   |   | 1 | 3 |   | 2 |   |   |   |   |   |   |   |   |   |   |   |
| p2  |   |   |   |   |   | 3 | 1 |   | 2 |   |   |   |   |   |   |   |   |   |   |   |
| p5  |   |   |   |   | 6 | 2 |   | 1 | 3 |   |   |   |   | 5 | 4 |   |   |   |   |   |
| p11 |   |   |   |   |   |   |   |   |   | 2 | 1 | 3 | 4 |   |   |   |   |   |   |   |
| p12 |   |   |   |   |   |   |   |   |   | 1 | 2 |   | 3 |   |   |   |   |   |   |   |
| p13 |   |   |   |   |   |   |   |   |   | 3 | 2 | 1 | 4 |   |   |   |   |   |   |   |
| p18 |   |   |   |   |   |   |   |   |   |   |   |   |   |   | 2 | 1 | 3 |   |   |   |
| p19 |   |   |   |   |   |   |   |   |   |   |   |   |   |   | 1 | 2 | 3 |   |   |   |
| p20 |   |   |   |   |   |   |   |   |   |   |   |   |   |   | 2 |   | 1 |   |   |   |
| p14 |   |   |   |   |   |   |   |   |   |   |   |   |   |   | 2 |   |   | 1 | 3 | 4 |
| p15 |   |   |   |   |   |   |   |   |   |   |   |   |   |   | 1 |   | 3 |   |   | 2 |
| p16 |   |   |   |   |   |   |   |   |   |   |   |   |   |   | 1 |   |   | 2 | 3 |   |
| p17 |   |   |   |   |   |   |   |   |   |   |   |   |   |   |   |   |   | 3 | 2 | 1 |

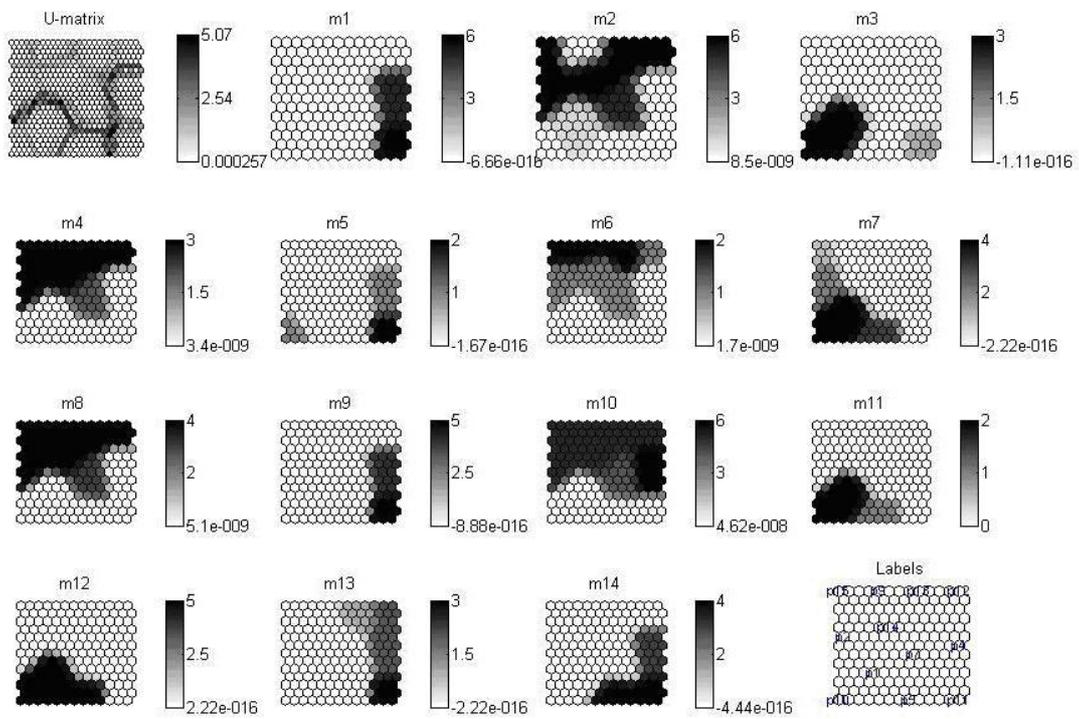

Figure 1. U-matrix, Component Plane and Labels of Self Organizing Map of Table 1 data

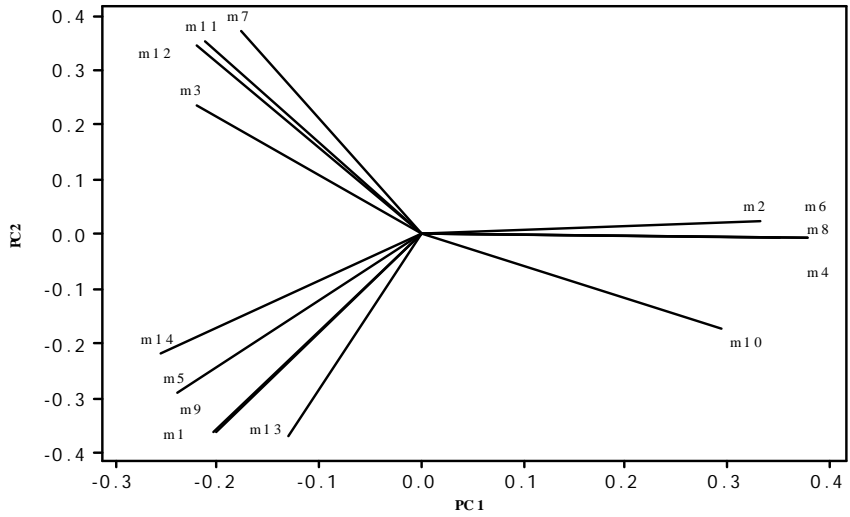

Figure 2a. Component loading plot of machine of Table 1 data

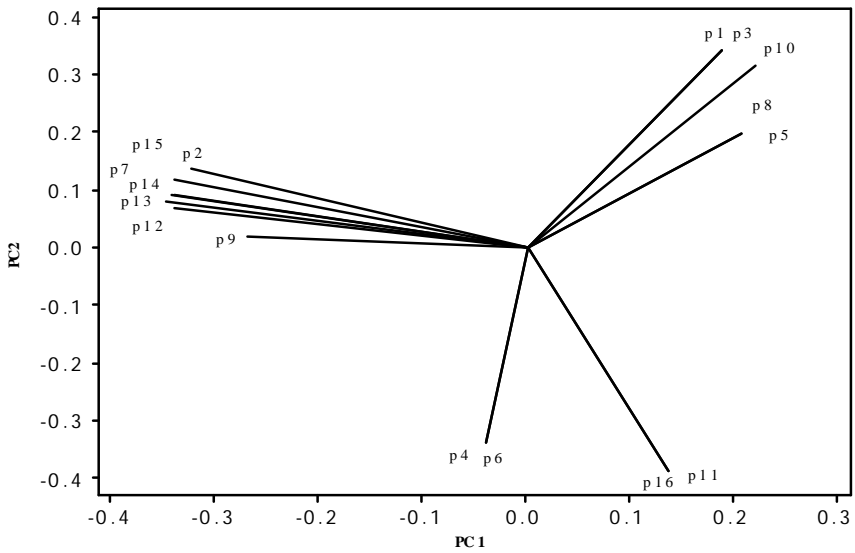

Figure 2b. Component loading plot of parts of Table 1 data

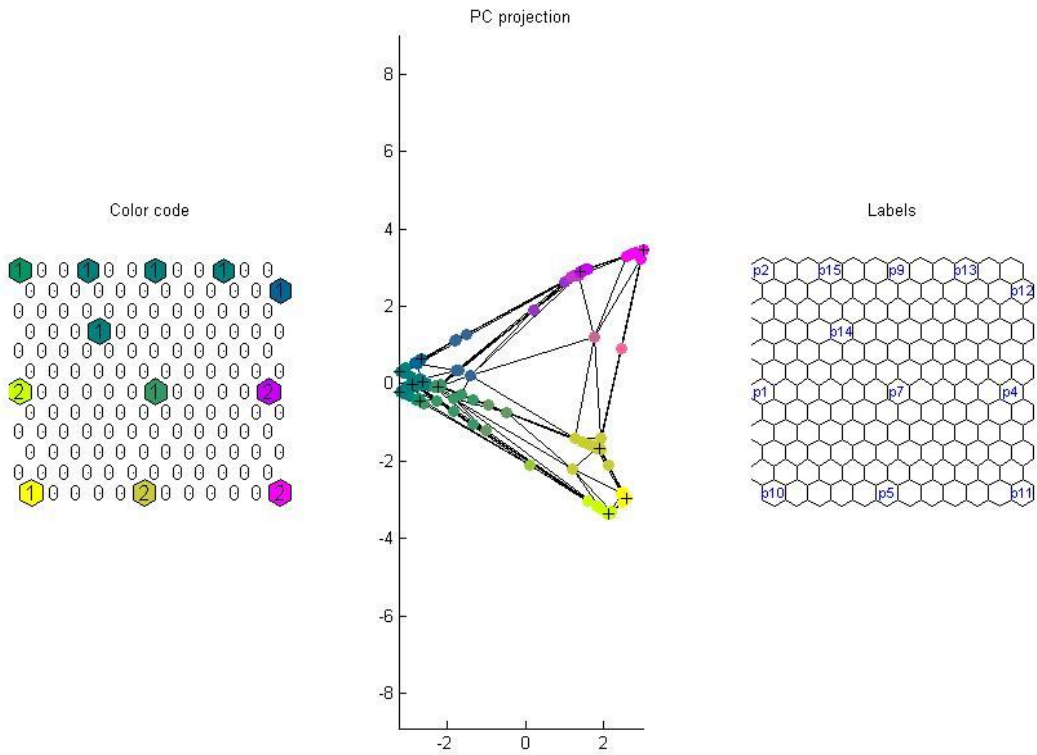

Figure 3. Extracted Color Code, PC projection and Labels in SOM Grid Map of Table 1 data

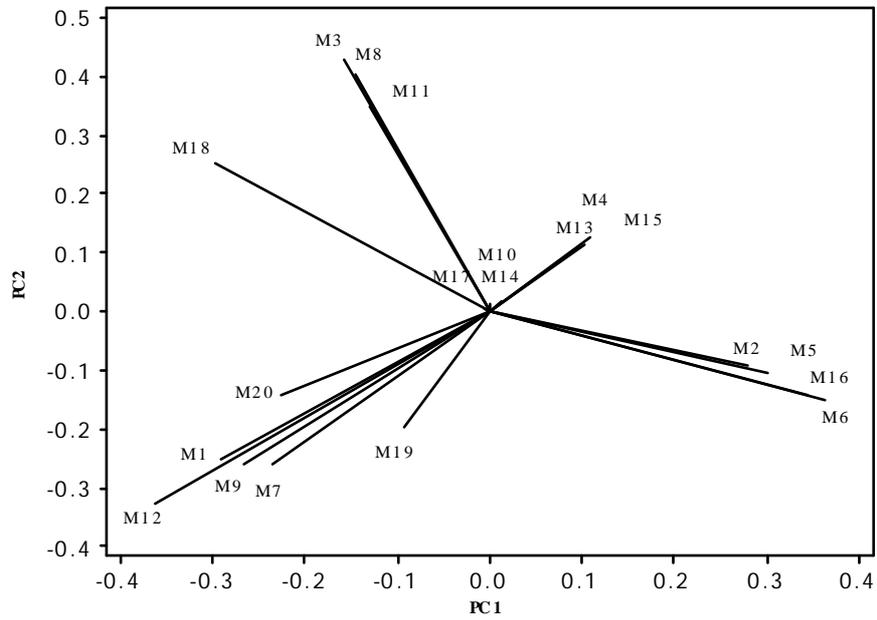

Figure 4a. Component loading plot of machines (Nagi et al., 1990)

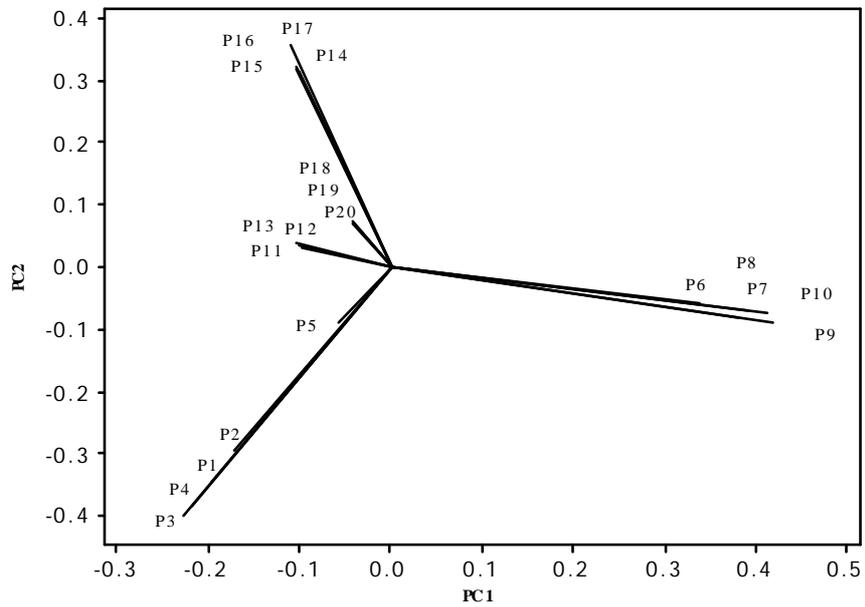

Figure 4b. Component loading plot of parts(Nagi et al., 1990)

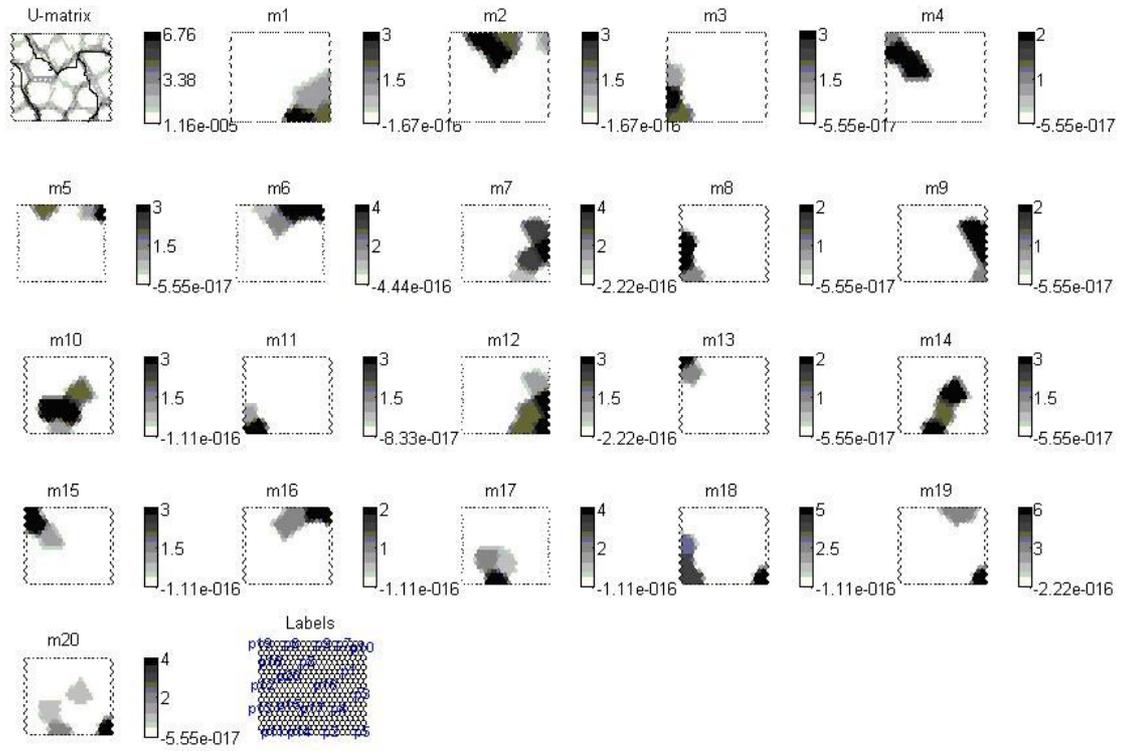

Figure 5. U-matrix, Component Plane and Labels of Self Organizing Map of problem from Nagi et al. (1990)

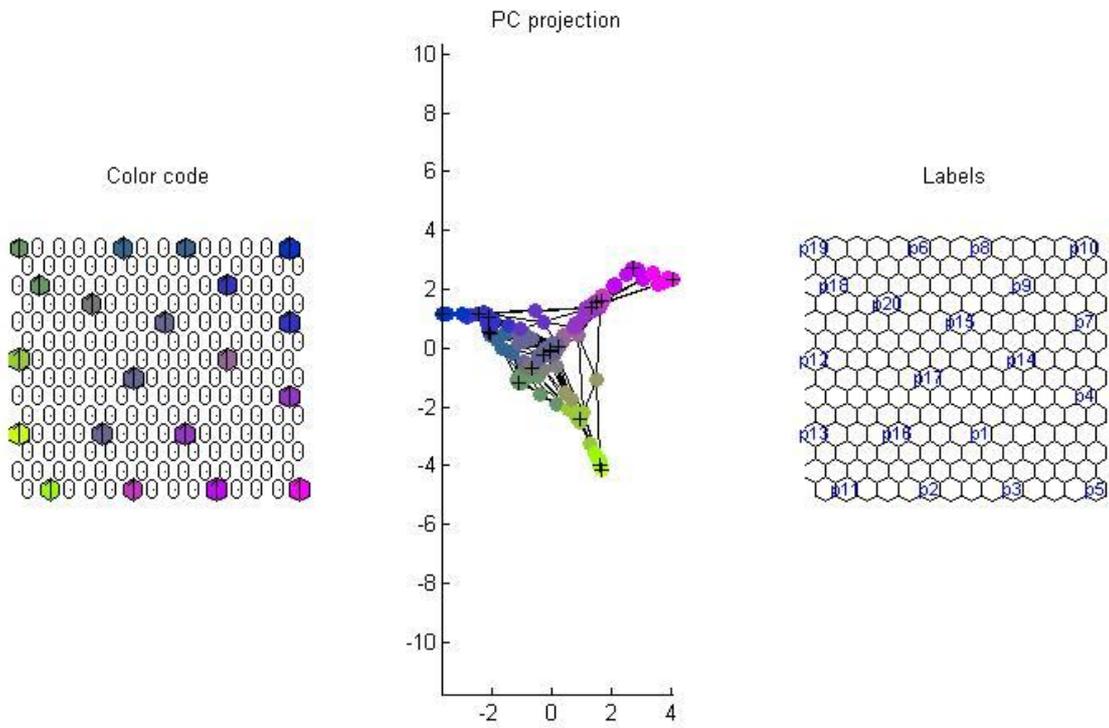

Figure 6. Extracted Color Code, PC projection and Labels in SOM Grid Map of dataset from Nagi et al., 1990